\documentclass{article}
\usepackage{spconf,amsmath,graphicx}
\usepackage[numbers,sort&compress]{natbib}
\usepackage{lineno,hyperref}
\usepackage{amssymb,amsfonts,amstext,graphicx}
\usepackage{epstopdf}
\usepackage{enumitem}
\usepackage{psfrag}
\usepackage{array}
\usepackage{epstopdf}
\usepackage{booktabs}
\usepackage{verbatim}
\usepackage{cases}
\usepackage{MnSymbol}
\usepackage{float}
\usepackage{longtable}
\usepackage[center]{caption2}
\allowdisplaybreaks
\newcommand{\vecnm}[1]{\left\|#1\right\|}

\title{Maximum Likelihood and Maximum A Posteriori Direction-of-Arrival Estimation in the Presence of SIRP Noise}
\name{Xin Zhang$^*$, Mohammed Nabil El Korso$^{**}$ and Marius Pesavento$^*$}
\address{$^*$Communication Systems Group, Technische Universit\"{a}t Darmstadt, Darmstadt, Germany\\
$^{**}$LEME EA416, Universit\'{e} Paris Ouest Nanterre La D\'{e}fense, Ville d{'}Avray, France}

\begin{document}
\ninept
\maketitle
\begin{abstract}
The maximum likelihood (ML) and maximum a posteriori (MAP) estimation techniques are widely used to address the direction-of-arrival (DOA) estimation problems, an important topic in sensor array processing. Conventionally the ML estimators in the DOA estimation context assume the sensor noise to follow a Gaussian distribution. In real-life application, however, this assumption is sometimes not valid, and it is often more accurate to model the noise as a non-Gaussian process. In this paper we derive an iterative ML as well as an iterative MAP estimation algorithm for the DOA estimation problem under the spherically invariant random process noise assumption, one of the most popular non-Gaussian models, especially in the radar context. Numerical simulation results are provided to assess our proposed algorithms and to show their advantage in terms of performance over the conventional ML algorithm.
\end{abstract}
\section{Introduction}\label{intro}
The direction-of-arrival (DOA) estimation problem is an important topic in sensor array processing which has found wide application in, among others, radar, sonar, radio astronomy and wireless communications \cite{HPRE14,zhxtc10,SN89,PG01}. Among the numerous techniques developed for the DOA estimation, those based on the maximum likelihood (ML) criterion are known to have the advantage of offering an outstanding tradeoff between the asymptotic and threshold performances \cite{SN89,SN90b}. Conventionally, a crucial assumption for the ML estimators is that the noise is \emph{uniformly white} \cite{SN89,SN90b}. Nevertheless, this oversimplifying assumption is unrealistic in certain applications \cite{GS02,wong1,ye1}. Thus, the authors of \cite{PG01} and \cite{vorobyov1,ollier1} have devised, resorting to the concept of \emph{stepwise numerical concentration}, an iterative ML estimator for the case of \emph{nonuniform white} and \emph{colored} noise, respectively.

The problem, however, is that the Gaussian noise assumption itself, colored or not, is based on the central limit theorem, and  loses immediately its validity in certain scenarios when the conditions for this are not fulfilled. This is the case, e.g., in the context of low-grazing-angle and/or high-resolution radar \cite{mim4,mim5,xzh3}, where the radar clutter shows non-stationarity. Various non-Gaussian noise models have been developed to deal with such problems, among which the so-called spherically invariant random process (SIRP) model has become the most notable and popular one \cite{mim6,mim5,mim8,mim9}. A SIRP is a two-scale, compound Gaussian process, formulated as the product of two components: the square root of a positive scalar random process, namely, the \textit{texture}, accounting for the local power changing, and a complex Gaussian process, namely, the \textit{speckle}, describing the local scattering. A SIRP is fully characterized by its texture parameter(s) and speckle covariance matrix.

The existing works addressing the estimation problems in a SIRP context almost exclusively assume the presence of secondary data (known noise-only realizations) in order to estimate the speckle and texture's parameters \cite{mim14,mim15,mim17,lombardo1,mim8,elkorso3,ollila1}, instead of unknown noise realizations embedded in and contaminating the received signal. In \cite{mim15}, the authors provided a parameter-expanded expectation-maximization (PX-EM) algorithm to estimate the unknown signal parameters under the SIRP noise. The problem they consider, however, is a linear one. Furthermore, the application of their algorithm is restricted to a special model, namely, the so-called generalized multivariate analysis of variance model \cite{dog1}. To the best of our knowledge, there are no algorithms available in the current literature for DOA estimation (a highly non-linear problem), nor for signal parameter estimation \emph{in general} in a comprehensive manner, under the SIRP noise. To fill this gap, and employing a similar methodology as in \cite{PG01} and \cite{vorobyov1}, we devise in this paper an iterative maximum likelihood estimation (IMLE) algorithm, together with an iterative maximum a posteriori estimation (IMAPE) algorithm in this context. The latter exploits information of the noise distribution and can be seen as a generalization of the former. Finally, we carry out simulation to illustrate the performances of our algorithms.

\section{Model Setup}\label{model}
Consider an arbitrary sensor array comprising $N$ sensors that receive $M$ ($M<N$) narrowband far-field source signals with unknown DOAs $\theta_1,\dots,\theta_M$. The array output at the $t$th snapshot can be formulated as \cite{SN89,SN90b}:
\begin{equation}\label{output}
\boldsymbol{x}(t)=\boldsymbol{A}\left(\boldsymbol{\theta}\right)\boldsymbol{s}(t)+\boldsymbol{n}(t),\quad t=1,\dots,T,
\end{equation}in which $\boldsymbol{\theta}=\left[\theta_1,\dots,\theta_M\right]^T$ is the $M\times1$ vector of unknown signal DOAs, $\boldsymbol{A}\left(\boldsymbol{\theta}\right)=\left[\boldsymbol{a}\left(\theta_1\right),\dots,\boldsymbol{a}\left(\theta_M\right)\right]$ denotes the $N\times M$ steering matrix, $\boldsymbol{s}(t)$ is the $M\times1$ vector of the source waveforms, $\boldsymbol{n}(t)$ is the $N\times1$ sensor noise vector, $T$ denotes the snapshot number, and $(\cdot)^T$ denotes transpose.

In this paper, we assume the source waveforms $\boldsymbol{s}(t),\ t=1,\dots,T$, to be unknown deterministic complex sequences \cite{SN89}. The sensor noise is modeled as a SIRP, which comprises two terms, statistically independent of each other \cite{mim6}:
\begin{equation}\label{5}
\boldsymbol{n}(t)=\sqrt{\tau(t)}\boldsymbol{\sigma}(t),\quad t=1,\dots,T;
\end{equation}in which $\boldsymbol{\sigma}(t)$ represents the speckle, a temporally white, complex Gaussian process with zero mean and an unknown $N\times N$ covariance matrix $\boldsymbol{Q}=\text{E}\left\{\boldsymbol{\sigma}(t)\boldsymbol{\sigma}^H(t)\right\}$, where $(\cdot)^{H}$ stands for the conjugate transpose; whereas the texture, denoted by $\tau(t)$, is composed of independent, identically distributed (i.i.d.) positive random variables at each snapshot. To resolve the ambiguity between the texture and the speckle so as to make the noise parameters uniquely identifiable, we assume that $\text{tr}\{\boldsymbol{Q}\}=N$, in which $\text{tr}\{\cdot\}$ denotes the trace. In this paper, we mainly consider two kinds of texture distributions that are most widely used in the literature, for both of which $\tau(t)$ is characterized by two parameters, the \emph{shape parameter} $a$ and the \emph{scale parameter} $b$. The first is the \emph{gamma distribution}, leading to the K-distributed noise \cite{nohara1,mim5}, where the pdf of $\tau(t)$ is:
\begin{equation}
p(\tau(t);a,b)=\frac{1}{\Gamma(a)b^a}\tau(t)^{a-1}e^{-\frac{\tau(t)}{b}},
\end{equation}in which $\Gamma(\cdot)$ denotes the gamma function. The second kind of our considered texture distribution is the \emph{inverse gamma distribution}, leading to the t-distributed noise \cite{lange1,jay2}, for which
\begin{equation}
p(\tau(t);a,b)=\frac{b^a}{\Gamma(a)}\tau(t)^{-a-1}e^{-\frac{b}{\tau(t)}}.
\end{equation}

Under the assumptions above, the unknown parameter vector of our problem is $\boldsymbol{\xi}=\left[\boldsymbol{\theta}^T,\boldsymbol{\chi}^T,\boldsymbol{\zeta}^T,a,b\right]^T$,
where $\boldsymbol{\chi}$ is a $2NT$-element vector containing the real and imaginary parts of the elements of $\boldsymbol{s}(t),\ t=1,\dots,T$, and $\boldsymbol{\zeta}$ is a $N^2$-element vector containing the real and imaginary parts of the entries of the lower triangular part of $\boldsymbol{Q}$.

Let $\boldsymbol{x}=\left[\boldsymbol{x}^T(1),...,\boldsymbol{x}^T(T)\right]^T$ denote the full observation vector, and $\boldsymbol{\tau}=\left[\tau(1),\dots,\tau(T)\right]^T$ represent the vector of texture realizations at all snapshots. The full observation likelihood conditioned on $\boldsymbol{\tau}$ can be written as:
\begin{equation}\label{n1}
p\left(\boldsymbol{x} | \boldsymbol{\tau}; \boldsymbol{\theta},\boldsymbol{\chi},\boldsymbol{\zeta} \right)=\prod_{t=1}^T\frac{\exp\left(-\frac{1}{\tau(t)}\boldsymbol{\rho}^H(t)\boldsymbol{\rho}(t)\right)}
{\mid\pi\tau(t)\boldsymbol{Q}\mid};
\end{equation}in which $\boldsymbol{\rho}(t)=
\boldsymbol{Q}^{-1/2}\left(\boldsymbol{x}(t)-\boldsymbol{A}\left(\boldsymbol{\theta}\right)\boldsymbol{s}(t)\right)$, represents the noise realization at snapshot $t$ with its speckle spatially whitened.

Eq.~(\ref{n1}), multiplied by $p(\boldsymbol \tau;a,b)$, leads to the joint likelihood of $\boldsymbol{x}$ and $\boldsymbol{\tau}$:
\begin{equation}\label{n1a}
\begin{aligned}
&p\left(\boldsymbol{x},  \boldsymbol{\tau}; \boldsymbol{\xi} \right)
=p\left(\boldsymbol{x} | \boldsymbol{\tau}; \boldsymbol{\theta},\boldsymbol{\chi},\boldsymbol{\zeta} \right)p(\boldsymbol \tau;a,b)\\
=&\prod_{t=1}^T\frac{\exp\left(-\frac{1}{\tau(t)}
\boldsymbol{\rho}^H(t)\boldsymbol{\rho}(t)\right)}{\mid\pi\tau(t)\boldsymbol{Q}\mid}p(\tau(t);a,b).
\end{aligned}
\end{equation}

\section{Iterative Maximum Likelihood Estimation}\label{sec3}
In our IMLE algorithm we maximize, similarly as in \cite{conte2}, the conditional likelihood in Eq.~(\ref{n1}), instead of the intractable marginal likelihood function, $\int_0^{+\infty}p\left(\boldsymbol{x},  \boldsymbol{\tau}; \boldsymbol{\xi} \right)\text{d}{\boldsymbol \tau}$, which does not yield a closed-form expression. In doing so, we actually focus on the texture realization $\boldsymbol{\tau}$, which is considered as deterministic, rather than the texture process itself.

%\subsection{Conditional Log-likelihood Function $\&$ Score Functions}
Let $L_\text{C}$ denote the conditional log-likelihood (LL) function, which arises from Eq.~(\ref{n1}), as:
\begin{equation} \label{ll_con}
\begin{aligned}
L_\text{C}=&
\ln p\left(\boldsymbol{x} | \boldsymbol{\tau}; \boldsymbol{\theta},\boldsymbol{\chi},\boldsymbol{\zeta} \right)
=-TN\ln\pi-T\ln|\boldsymbol{Q}|\\
&-N\sum_{t=1}^{T}\ln\tau(t)
-\sum_{t=1}^{T}\frac{1}{\tau(t)}\boldsymbol{\rho}^H(t)\boldsymbol{\rho}(t).
\end{aligned}
\end{equation}

To begin with, we set $\partial L_\text{C}/\partial\tau(t)=0$, the solution of which provides an estimate of the parameter $\tau(t)$ when the parameters $\boldsymbol{\theta}$, $\boldsymbol{s}(t)$ and $\boldsymbol{Q}$ are fixed. We denote this estimate by $\hat{\tau}(t)$, which has the following expression:
\begin{equation}\label{tau_es}
\begin{aligned}
\hat{\tau}(t)
=\frac{1}{N}\left(\boldsymbol{x}(t)-\boldsymbol{A}\left(\boldsymbol{\theta}\right)\boldsymbol{s}(t)\right)^H\boldsymbol{Q}^{-1}
\left(\boldsymbol{x}(t)-\boldsymbol{A}\left(\boldsymbol{\theta}\right)\boldsymbol{s}(t)\right).
\end{aligned}
\end{equation}
Meanwhile, by applying Lemma 3.2.2. in \cite{And84} to Eq.~(\ref{ll_con}), one can obtain the expression of $\hat{\boldsymbol{Q}}$, representing the estimate of $\boldsymbol{Q}$ when $\boldsymbol{\theta}$, $\boldsymbol{s}(t)$ and $\tau(t)$ and are fixed, as:
\begin{equation}\label{Sig_es}
\hat{\boldsymbol{Q}}=\frac{1}{T}\sum_{t=1}^{T}\frac{1}{\tau(t)}
\left(\boldsymbol{x}(t)-\boldsymbol{A}\left(\boldsymbol{\theta}\right)\boldsymbol{s}(t)\right)
\left(\boldsymbol{x}(t)-\boldsymbol{A}\left(\boldsymbol{\theta}\right)\boldsymbol{s}(t)\right)^H,
\end{equation}in which replacing $\tau(t)$ by the expression of $\hat{\tau}(t)$ in Eq.~(\ref{tau_es}) leads to the following iterative expression of $\hat{\boldsymbol{Q}}$:
\begin{equation}\label{Sig_es2}
\hat{\boldsymbol{Q}}^{(i+1)}=\frac{N}{T}\sum_{t=1}^{T}
\frac{\left(\boldsymbol{x}(t)-\boldsymbol{A}\left(\boldsymbol{\theta}\right)\boldsymbol{s}(t)\right)\left(\boldsymbol{x}(t)-\boldsymbol{A}\left(\boldsymbol{\theta}\right)\boldsymbol{s}(t)\right)^H}
{\left(\boldsymbol{x}(t)-\boldsymbol{A}\left(\boldsymbol{\theta}\right)\boldsymbol{s}(t)\right)^H\left(\hat{\boldsymbol{Q}}^{(i)}\right)^{-1}
\left(\boldsymbol{x}(t)-\boldsymbol{A}\left(\boldsymbol{\theta}\right)\boldsymbol{s}(t)\right)},
\end{equation}for which we choose the identity matrix of size $N$, denoted by $\boldsymbol{I}_{N}$, to serve as the initialization matrix $\hat{\boldsymbol{Q}}^{(0)}$.

We further need to normalize $\hat{\boldsymbol{Q}}^{(i+1)}$ in Eq.~(\ref{Sig_es2}) to fulfill the assumption that $\text{tr}\{\boldsymbol{Q}\}=N$. Let $\hat{\boldsymbol{Q}}^{(i+1)}_\text{n}$ denote the normalized estimate $\hat{\boldsymbol{Q}}^{(i+1)}$, which is:
\begin{equation}\label{Sig_es_n}
\hat{\boldsymbol{Q}}^{(i+1)}_\text{n}=
N\left.\hat{\boldsymbol{Q}}^{(i+1)}\middle/\text{tr}\left\{\hat{\boldsymbol{Q}}^{(i+1)}\right\}\right..
\end{equation}

Now we consider the estimate of $\boldsymbol{s}(t)$ when $\boldsymbol{\theta}$, $\tau(t)$ and $\boldsymbol{Q}$ are fixed, which, denoted by $\hat{\boldsymbol{s}}(t)$, can be found by solving $\partial L_\text{C}/\partial \boldsymbol{s}(t)=0$, as:
\begin{equation}\label{s_es}
\hat{\boldsymbol{s}}(t)=\left(\tilde{\boldsymbol{A}}^H\left(\boldsymbol{\theta}\right)
\tilde{\boldsymbol{A}}\left(\boldsymbol{\theta}\right)\right)^{-1}
\tilde{\boldsymbol{A}}^H\left(\boldsymbol{\theta}\right)
\tilde{\boldsymbol{x}}(t),
\end{equation}in which $\tilde{\boldsymbol{A}}\left(\boldsymbol{\theta}\right)=\boldsymbol{Q}^{-1/2}\boldsymbol{A}\left(\boldsymbol{\theta}\right)$, $\tilde{\boldsymbol{x}}(t)=\boldsymbol{Q}^{-1/2}\boldsymbol{x}(t)$, representing the steering matrix and the observation at snapshot $t$, both pre-whitened by the speckle covariance matrix $\boldsymbol{Q}$, respectively.

From Eqs.~(\ref{tau_es}), (\ref{Sig_es2}) and (\ref{s_es}) one can see that the estimates of $\tau(t)$, $\boldsymbol{Q}$ and $\boldsymbol{s}(t)$ are mutually dependent, and further dependent on the parameter vector $\boldsymbol{\theta}$. This dependency makes it impossible to obtain a closed-form expression for the LL function concentrated w.r.t. each of the individual parameters $\tau(t)$, $\boldsymbol{Q}$ and $\boldsymbol{s}(t)$ and independent of other unknown parameters. To cope with this difficulty, we appeal to the so-called \emph{stepwise numerical concentration} method introduced in \cite{PG01,vorobyov1}, and concentrate the LL function iteratively. This can be accomplished by assuming at a particular iteration that, in our case, $\hat{\boldsymbol{Q}}$ and $\hat{\tau}(t)$ are known and can be used in the computation of $\hat{\boldsymbol{s}}(t)$, which is then used in its turn to update $\hat{\boldsymbol{Q}}$ and $\hat{\tau}(t)$ in the next iteration. The sequential updating procedure is repeated until convergence.

Finally, we address the estimation of $\boldsymbol{\theta}$, our parameter of interest, considering the values of $\boldsymbol{Q}$ and $\boldsymbol{\tau}$ as fixed and known. Thus, neglecting the constant terms, the conditional LL function in Eq.~(\ref{ll_con}) can be reformulated as:
\begin{equation} \label{ll_con2}
L_\text{C}=
-\sum_{t=1}^{T}\frac{1}{\tau(t)}\boldsymbol{\rho}^H(t)\boldsymbol{\rho}(t),
\end{equation}into which we insert Eq.~(\ref{s_es}). The resulting expression is then maximized w.r.t. $\boldsymbol{\theta}$, to obtain the estimate of $\boldsymbol{\theta}$ for each iteration, denoted by $\hat{\boldsymbol{\theta}}$, as:
\begin{equation}\label{theta_es}
\hat{\boldsymbol{\theta}}
=\arg \min_{\boldsymbol{\theta}}\left\{\sum_{t=1}^{T}\frac{1}{\tau(t)}
\vecnm{\boldsymbol{P}_{\tilde{\boldsymbol{A}}(\boldsymbol{\theta})}^{\bot}(t)\tilde{\boldsymbol{x}}(t)}^2\right\},
\end{equation}in which $\vecnm{\cdot}$ denotes the Euclidean norm and $\boldsymbol{P}_{\tilde{\boldsymbol{A}}(\boldsymbol{\theta})}^{\bot}(t)=\boldsymbol{I}_{N}
-\tilde{\boldsymbol{A}}(\boldsymbol{\theta})\left(\tilde{\boldsymbol{A}}^H(\boldsymbol{\theta})
\tilde{\boldsymbol{A}}(\boldsymbol{\theta})\right)^{-1}\tilde{\boldsymbol{A}}^H(\boldsymbol{\theta})$, stands for the orthogonal projection matrix onto the null space of the matrix $\tilde{\boldsymbol{A}}(\boldsymbol{\theta})$.

Our proposed IMLE algorithm, comprising three steps, can be summarized as follows:

\noindent\textbf{Step 1}: Initialization. At iteration $i=0$, set $\hat{\tau}^{(0)}(t)=1,\ t=1,\dots,T$, and $\hat{\boldsymbol{Q}}^{(0)}_\text{n}=\boldsymbol{I}_N$.

\noindent\textbf{Step 2}: Calculate $\hat{\boldsymbol{\theta}}^{(i)}$ from Eq.~(\ref{theta_es}) using $\hat{\tau}^{(i)}(t)$ and $\hat{\boldsymbol{Q}}^{(i)}_\text{n}$, then $\hat{\boldsymbol{s}}^{(i)}(t)$ from Eq.~(\ref{s_es}) using $\hat{\boldsymbol{\theta}}^{(i)}$, $\hat{\tau}^{(i)}(t)$ and $\hat{\boldsymbol{Q}}^{(i)}_\text{n}$.

\noindent\textbf{Step 3}: Use $\hat{\boldsymbol{\theta}}^{(i)}$, $\hat{\boldsymbol{s}}^{(i)}(t)$ and $\hat{\boldsymbol{Q}}^{(i)}_\text{n}$ to update $\hat{\boldsymbol{Q}}^{(i+1)}_\text{n}$ from Eqs.~(\ref{Sig_es2}) and (\ref{Sig_es_n}). Then use $\hat{\boldsymbol{\theta}}^{(i)}$, $\hat{\boldsymbol{s}}^{(i)}(t)$ and the updated matrix $\hat{\boldsymbol{Q}}^{(i+1)}_\text{n}$ to find the update $\hat{\tau}^{(i+1)}(t)$ from Eq.~(\ref{tau_es}). Set $i=i+1$.

Repeat Step 2 and Step 3 until a stop criterion (convergence or a maximum number of iteration) to obtain the final estimate of $\boldsymbol{\theta}$, denoted by $\hat{\boldsymbol{\theta}}_\text{IMLE}$.

The convergence of our algorithm is guaranteed by the fact that the value of the objective function in Eq.~(\ref{theta_es}) at each step can either improve or maintain but cannot worsen \cite{vorobyov1}. The same holds true for the update of $\hat{\boldsymbol{Q}}$ and $\hat{\tau}(t)$. In fact, as our simulations will show, the convergence can be attained by only two iterations. Thus, the computational cost of our algorithm, which lies mainly in the solution of the highly nonlinear optimization problem in Step 2, is only a few times of that of the conventional ML estimation (CMLE) algorithm, which, incidentally, corresponds to the case in which the noise is uniform white Gaussian, such that Eq.~(\ref{theta_es}) degenerates into:
\begin{equation}\label{conv_ml}
\hat{\boldsymbol{\theta}}_\text{CMLE}= \arg\min_{\boldsymbol{\theta}}\left\{\sum_{t=1}^{T}\vecnm{\boldsymbol{P}_{{\boldsymbol{A}}
(\boldsymbol{\theta})}^{\bot}{\boldsymbol{x}}(t)}^2\right\}.
\end{equation}

\section{Iterative Maximum A Posteriori Estimation}\label{sec3b}
The IMLE algorithm, presented in Section \ref{sec3}, treats the texture as deterministic and thereby ignores information of its statistical properties. This has the advantage of easier and faster implementation, and is also a natural approach when the texture distribution is either unknown or does not have a closed-form expression, e.g., in the case of Weibull-distributed noise. In general cases, however, such approach is \emph{suboptimal}. Thus, when the texture distribution is available, we have the better choice of exploiting information from the texture's prior distribution, i.e., employing the maximum a posteriori (MAP) approach, in designing our estimation procedure. This leads to our IMAPE algorithm that we propose in this section.

The MAP estimator maximizes the joint LL function, denoted by $L_\text{J}$, which is equal to:
\begin{equation} \label{ll_joi}
\begin{aligned}
L_\text{J}&=\ln p\left(\boldsymbol{x},  \boldsymbol{\tau}; \boldsymbol{\xi} \right)
=
\ln \left(p\left(\boldsymbol{x} | \boldsymbol{\tau}; \boldsymbol{\theta},\boldsymbol{\chi},\boldsymbol{\zeta} \right)p(\boldsymbol \tau;a,b)\right)\\
&=L_\text{C}+\sum_{t=1}^T \ln p(\tau(t);a,b)\\
&=\left\{
\begin{aligned}
&L_\text{C}-T\ln\Gamma(a)-Ta\ln{b}+(a-1)\sum_{t=1}^{T}\ln\tau(t)\\
&-\frac{\sum_{t=1}^{T}\tau(t)}{b}, \quad \text{K-distributed noise},\\
&L_\text{C}-T\ln\Gamma(a)+Ta\ln{b}-(a+1)\sum_{t=1}^{T}\ln\tau(t)\\
&-b\sum_{t=1}^{T}\frac{1}{\tau(t)}, \quad \text{t-distributed noise}.
\end{aligned}
\right.
\end{aligned}
\end{equation}

Solving $\partial L_\text{J}/\partial\tau(t)=0$ leads to the expression of $\hat{\tau}(t)$ when all the remaining unknown parameters are fixed, which is:
\begin{equation}\label{tau_es2}
\hat{\tau}(t)=\left\{
\begin{aligned}
&\frac{1}{2}\bigg(\left(a-N-1\right)b+\Big(\left(a-N-1\right)^2b^2\\
&+4b
\left(\boldsymbol{x}(t)-\boldsymbol{A}\left(\boldsymbol{\theta}\right)\boldsymbol{s}(t)\right)^H\boldsymbol{Q}^{-1}\\
&\cdot\left(\boldsymbol{x}(t)-\boldsymbol{A}\left(\boldsymbol{\theta}\right)\boldsymbol{s}(t)\right)\Big)^\frac{1}{2}\bigg),\quad  \text{K-distributed noise},\\
&\frac{1}{a+N+1}\big(\left(\boldsymbol{x}(t)-\boldsymbol{A}\left(\boldsymbol{\theta}\right)
\boldsymbol{s}(t)\right)^H\boldsymbol{Q}^{-1}\\
&\cdot\left(\boldsymbol{x}(t)-\boldsymbol{A}\left(\boldsymbol{\theta}\right)\boldsymbol{s}(t)\right)+b\big),
\quad \text{t-distributed noise}.
\end{aligned}
\right.
\end{equation}

Next we consider the estimation of the texture parameters $a$ and $b$, denoted by $\hat{a}$ and $\hat{b}$. The latter can be obtained by solving $\partial L_\text{J}/\partial b=0$, as:
\begin{equation}\label{b_es}
\hat{b}=\left\{
\begin{aligned}
&\frac{\sum_{t=1}^{T}\tau(t)}{Ta}, \quad \text{K-distributed noise},\\
&\frac{Ta}{\sum_{t=1}^{T}\frac{1}{\tau(t)}}, \quad \text{t-distributed noise}.
\end{aligned}
\right.
\end{equation}Meanwhile, calculation of $\partial L_\text{J}/\partial a$ results in:
\begin{equation}\label{a_es}
\frac{\partial L_\text{J}}{\partial a}=\left\{
\begin{aligned}
&-T\Psi(a)-T\ln b+\sum_{t=1}^{T}\ln\tau(t), \ \text{K-distributed noise},\\
&-T\Psi(a)+T\ln b-\sum_{t=1}^{T}\ln\tau(t), \ \text{t-distributed noise};
\end{aligned}
\right.
\end{equation}in which $\Psi(\cdot)$ stands for the digamma function. It is obvious from Eq.~(\ref{a_es}) that $\partial L_\text{J}/\partial a=0$ does not allow an analytical expression of the root. Thus $\hat{a}$ can only be calculated numerically.

Next, we approach the estimation of the source waveforms and the speckle covariance matrix. By noticing that $\partial L_\text{J}/\partial {\boldsymbol{Q}}=\partial L_\text{C}/\partial {\boldsymbol{Q}}$, and $\partial L_\text{J}/\partial {\boldsymbol{s}(t)}=\partial L_\text{C}/\partial {\boldsymbol{s}(t)}$, it follows immediately that the same expressions of $\hat{\boldsymbol{Q}}$ and $\hat{\boldsymbol{s}}(t)$ in Eqs.~(\ref{Sig_es}) and (\ref{s_es}), which we obtained for the IMLE algorithm, are also valid in the case of the IMAPE algorithm. Substituting into Eq.~(\ref{Sig_es}) the new expression of $\hat{\tau}(t)$ in Eq.~(\ref{tau_es2}) leads to the following expression for $\hat{\boldsymbol{Q}}$:
\begin{equation}\label{Sig_es4}
\hat{\boldsymbol{Q}}^{(i+1)}=\left\{
\begin{aligned}
&\frac{2}{T}\sum_{t=1}^{T}\left(\boldsymbol{x}(t)-\boldsymbol{A}\left(\boldsymbol{\theta}\right)\boldsymbol{s}(t)\right)\\
&\cdot\left(\boldsymbol{x}(t)-\boldsymbol{A}\left(\boldsymbol{\theta}\right)\boldsymbol{s}(t)\right)^H
\\&
\left.\middle/\Bigg(\right.\bigg(4b
\left(\boldsymbol{x}(t)-\boldsymbol{A}\left(\boldsymbol{\theta}\right)\boldsymbol{s}(t)\right)^H
\left(\hat{\boldsymbol{Q}}^{(i)}\right)^{-1}\\
&\cdot\left(\boldsymbol{x}(t)-\boldsymbol{A}\left(\boldsymbol{\theta}\right)\boldsymbol{s}(t)\right)
+\left(a-N-1\right)^2b^2
\bigg)^\frac{1}{2}\\&
+\left(a-N-1\right)b\Bigg),\quad  \text{K-distributed noise},\\
&\frac{a+N+1}{T}\sum_{t=1}^{T}
\Big(\left(\boldsymbol{x}(t)-\boldsymbol{A}\left(\boldsymbol{\theta}\right)\boldsymbol{s}(t)\right)\\
&\cdot\left(\boldsymbol{x}(t)-\boldsymbol{A}\left(\boldsymbol{\theta}\right)\boldsymbol{s}(t)\right)^H\Big)\\
&\left.\middle/\bigg(\right.b+\left(\boldsymbol{x}(t)-\boldsymbol{A}\left(\boldsymbol{\theta}\right)\boldsymbol{s}(t)\right)^H
\left(\hat{\boldsymbol{Q}}^{(i)}\right)^{-1}\\
&\cdot\left(\boldsymbol{x}(t)-\boldsymbol{A}\left(\boldsymbol{\theta}\right)\boldsymbol{s}(t)\right)\bigg),\quad \text{t-distributed noise}.
\end{aligned}
\right.
\end{equation}which, similar to the expression of $\hat{\boldsymbol{Q}}^{(i+1)}$ in Eq.~(\ref{Sig_es2}) for the IMLE algorithm, needs to be substituted into Eq.~(\ref{Sig_es_n}) to obtain the normalized $\hat{\boldsymbol{Q}}^{(i+1)}$ denoted as $\hat{\boldsymbol{Q}}^{(i+1)}_\text{n}$.

Finally, we address the estimation of $\boldsymbol{\theta}$. Adopting the numerical concentration method similar to that in Section \ref{sec3}, we also assume here that $\boldsymbol{Q}$ and $\boldsymbol{\tau}$ are known from the previous iteration of the algorithm. Furthermore, as the estimates of $a$ and $b$ are only dependent on $\boldsymbol{\tau}$, these are also fixed for each iteration. This allows us to drop those terms in the expression of the joint LL function $L_\text{J}$ in Eq.~(\ref{ll_joi}) that contain only these unknown parameters, and thereby to transform it into the same expression as in Eq.~(\ref{ll_con2}). This means that $\boldsymbol{\theta}$ can be obtained, also for the IMAPE algorithm, from Eq.~(\ref{theta_es}).

The iterative estimation procedure of our IMAPE algorithm also contains three steps, and is summarized as follows:

\noindent\textbf{Step 1}: Initialization. At iteration $i=0$, set $\hat{\tau}^{(0)}(t),\ t=1,\dots,T$ as the absolute values of independent random numbers from the standard normal distribution\footnote{Unlike the case of the IMLE algorithm, for the IMAPE algorithm, which involves estimation of the texture parameters, initializing the texture components as all ones will lead to poor performance. We thus initialize here the texture as random numbers instead.}, and $\hat{\boldsymbol{Q}}^{(0)}_\text{n}=\boldsymbol{I}_N$.

\noindent\textbf{Step 2}: Calculate $\hat{\boldsymbol{\theta}}^{(i)}$ from Eq.~(\ref{theta_es}) using $\hat{\tau}^{(i)}(t)$ and $\hat{\boldsymbol{Q}}^{(i)}_\text{n}$, then $\hat{\boldsymbol{s}}^{(i)}(t)$ from Eq.~(\ref{s_es}) using $\hat{\boldsymbol{\theta}}^{(i)}$, $\hat{\tau}^{(i)}(t)$ and $\hat{\boldsymbol{Q}}^{(i)}_\text{n}$. Meanwhile, substitute Eq.~(\ref{b_es}) into Eq.~(\ref{a_es}). First find numerically $\hat{a}^{(i)}$ from Eq.~(\ref{a_es}) using $\hat{\tau}^{(i)}(t)$, then $\hat{b}^{(i)}$ from Eq.~(\ref{b_es}) using $\hat{\tau}^{(i)}(t)$ and $\hat{a}^{(i)}$.

\noindent\textbf{Step 3}: Use $\hat{\boldsymbol{\theta}}^{(i)}$, $\hat{\boldsymbol{s}}^{(i)}(t)$, $\hat{\boldsymbol{Q}}^{(i)}_\text{n}$, $\hat{a}^{(i)}$ and $\hat{b}^{(i)}$ to update $\hat{\boldsymbol{Q}}^{(i+1)}_\text{n}$ from Eqs.~(\ref{Sig_es4}) and (\ref{Sig_es_n}). Then use $\hat{\boldsymbol{\theta}}^{(i)}$, $\hat{\boldsymbol{s}}^{(i)}(t)$, $\hat{a}^{(i)}$, $\hat{b}^{(i)}$ and the updated matrix $\hat{\boldsymbol{Q}}^{(i+1)}_\text{n}$ to find the update $\hat{\tau}^{(i+1)}(t)$ from Eq.~(\ref{tau_es2}). Set $i=i+1$.

Repeat Step 2 and Step 3 until a stop criterion (convergence or a maximum number of iteration) to obtain the final $\hat{\boldsymbol{\theta}}$, denoted by $\hat{\boldsymbol{\theta}}_\text{IMAPE}$.

The remarks at the end of Section \ref{sec3}, upon the convergence and computational cost of our IMLE algorithm, also directly apply to our IMAPE algorithm.

\section{Numerical Simulations}
In our simulations we consider a uniform linear array comprising $N=6$ omnidirectional sensors with half-wavelength inter-sensor spacing. Two equally powered narrowband sources impinge on the array with the DOAs $\theta_1=30^{\circ}$ and $\theta_2=60^{\circ}$ relative to the broadside. The number of statistically independent snapshots is $T=10$. For K-distributed sensor noise we choose $a=1.6$ and $b=2$; and for t-distributed sensor noise, $a=1.1$ and $b=2$. The entries of the speckle covariance matrix $\boldsymbol{Q}$ are generated by \cite{VSO97} $[\boldsymbol{Q}]_{m,n}=\sigma^2\cdot0.9^{|m-n|}e^{j\frac{\pi}{2}(m-n)}, \ m,n=1,\dots,N$. The number of Monte-Carlo trials is $100$. The signal-to-noise ratio (SNR) is defined as:
\begin{equation}\label{snr}
\text{SNR}=\frac{\sum_{t=1}^{T}\vecnm{\boldsymbol{s}(t)}^2}
{T\text{E}\{\tau(t)\}\text{tr}\left\{\boldsymbol{Q}\right\}},
\end{equation}in which $\text{E}\{\tau(t)\}$ is equal to $ab$ for a K-distributed noise and $b/(a-1)$ for a t-distributed noise (for $a>1$) \cite{pp95}.

\begin{figure}[htpb]
  \centerline{\includegraphics[width=0.5\textwidth]{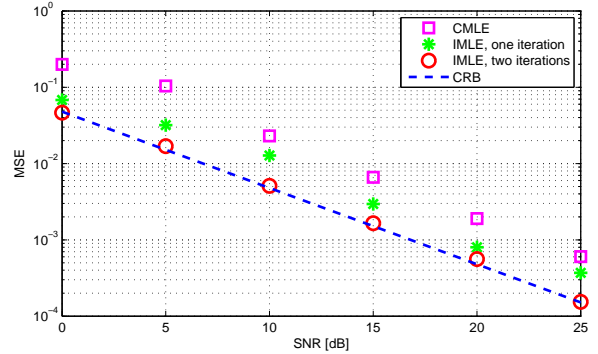}}
  \caption{MSE vs. SNR under t-distributed noise.}
  \label{fig1}
  \end{figure}

\begin{figure}[htpb]
    \centerline{\includegraphics[width=0.5\textwidth]{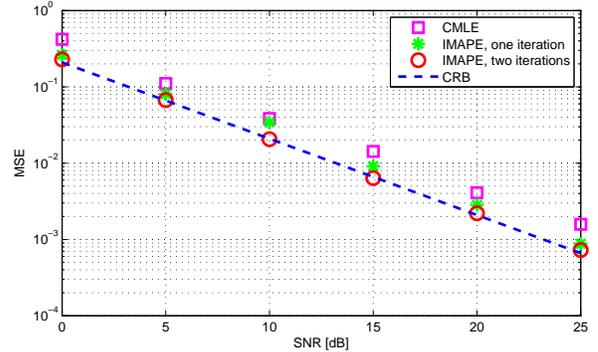}}
  \caption{MSE vs. SNR under K-distributed noise.}
  \label{fig2}
\end{figure}

In Figs.~\ref{fig1} and \ref{fig2}, we plot the mean square errors (MSEs) of the estimation of $\boldsymbol{\theta}$ under the SIRP noise versus the SNR by implementing our proposed IMLE and IMAPE algorithms, respectively. In Fig.~\ref{fig1} the noise is t-distributed, and in Fig.~\ref{fig2}, K-distributed. For comparison we also plot, in both figures, the MSEs generated by the CMLE algorithm in Eq.~(\ref{conv_ml}), and the deterministic Cram\'{e}r-Rao bound (CRB) \cite{xzh3}. From these figures one can clearly see that the conventional ML algorithm becomes poor under the SIRP noise, and both of our algorithms lead to significantly superior performance. These figures also show that only two iterations are sufficient for both of our algorithms to have a satisfactory performance, in terms of a resulting MSE appropriately close to $\text{CRB}(\boldsymbol{\theta})$, in asymptotic SNR cases.

\section{Conclusion}
In this paper we addressed the problem of estimating the DOAs of multiple sources under the SIRP noise, by deriving two new estimators belonging respectively to the ML and the MAP family. Our proposed IMLE and IMAPE algorithms are both based on the stepwise concentration of the LL function w.r.t. signal and noise parameters. As our simulations show, both algorithms require only a few iterations to attain convergence, and lead to significantly superior performance than the conventional approach.

\bibliographystyle{IEEEbib}
\bibliography{zhxtc}

\begin{thebibliography}{10}

\bibitem{HPRE14}
M.~Haardt, M.~Pesavento, F.~R\"{o}emer, and M.~N. {El~Korso},
\newblock {\em {Subspace Methods and Exploitation of Special Array Structures,
  Electronic Reference in Signal Processing: Array and Statistical Signal
  Processing (M. Viberg, ed.)}}, vol.~3,
\newblock Academic Press Library in Signal Processing, Elsevier Ltd., 2014.

\bibitem{zhxtc10}
H.~L.~Van Trees,
\newblock {\em Optimum Array Processing: {P}art {IV} of Detection, Estimation,
  and Modulation Theory},
\newblock John Wiley \& Sons Inc., New York, 2002.

\bibitem{SN89}
P.~Stoica and A.~Nehorai,
\newblock ``{MUSIC}, maximum likelihood and {C}ramer-{R}ao bound,''
\newblock {\em {IEEE} Trans. Acoust., Speech, Signal Processing}, vol. 37, no.
  5, pp. 720--741, May 1989.

\bibitem{PG01}
M.~Pesavento and A.~B. Gershman,
\newblock ``Maximum-likelihood direction-of-arrival estimation in the presence
  of unknown nonuniform noise,''
\newblock {\em {IEEE} Trans. Signal Processing}, vol. 49, no. 7, pp.
  1310--1324, July 2001.

\bibitem{SN90b}
P.~Stoica and A.~Nehorai,
\newblock ``Performances study of conditional and unconditional direction of
  arrival estimation,''
\newblock {\em {IEEE} Trans. Acoust., Speech, Signal Processing}, vol. 38, pp.
  1783--1795, Oct. 1990.

\bibitem{GS02}
A.B. Gershman, P.~Stoica, M.~Pesavento, and E.G. Larsson,
\newblock ``Stochastic {C}ram\'er-{R}ao bound for direction estimation in
  unknown noise fields,''
\newblock {\em IEE Proceedings-Radar, Sonar and Navigation}, vol. 149, pp.
  2--8, Jan. 2002.

\bibitem{wong1}
K.~M. Wong, J.~P. Reilly, Q.~Wu, and S.~Qiao,
\newblock ``Estimation of the directions of arrival of signals in unknown
  correlated noise. {P}art {I}: {T}he {MAP} approach and its implementation,''
\newblock {\em IEEE Trans. Signal Processing}, vol. 40, no. 8, pp. 2007--2017,
  Aug. 1992.

\bibitem{ye1}
H.~Ye and R.~D. DeGroat,
\newblock ``Maximum likelihood {DOA} estimation and asymptotic
  {C}ram\'{e}r-{R}ao bounds for additive unknown colored noise,''
\newblock {\em IEEE Trans. Signal Processing}, vol. 43, no. 4, pp. 938--949,
  Apr. 1995.

\bibitem{vorobyov1}
S.~A. Vorobyov, A.~B. Gershman, and K.~M. Wong,
\newblock ``Maximum likelihood direction-of-arrival estimation in unknown noise
  fields using sparse sensor arrays,''
\newblock {\em IEEE Trans. Signal Processing}, vol. 53, no. 1, pp. 34--43, Jan.
  2005.

\bibitem{ollier1}
V.~Ollier, M.~N. {El~Korso}, R.~Boyer, P.~Larzabal, and M.~Pesavento,
\newblock ``Joint {ML} calibration and {DOA} estimation with separated
  arrays,''
\newblock in {\em Proc. ICASSP}, Shanghai, China, Mar. 2016.

\bibitem{mim4}
J.~B. Billingsley,
\newblock ``Ground clutter measurements for surface-sited radar,''
\newblock Tech. {R}ep. 780, Massachusetts Inst. Technol., Cambridge, MA, Feb.
  1993.

\bibitem{mim5}
F.~Gini, M.~V. Greco, M.~Diani, and L.~Verrazzani,
\newblock ``Performance analysis of two adaptive radar detectors against
  non-{G}aussian real sea clutter data,''
\newblock {\em IEEE Trans. Aerosp. Electron. Syst.}, vol. 36, no. 4, pp.
  1429--1439, Oct. 2000.

\bibitem{xzh3}
X.~Zhang, M.~N. {El~Korso}, and M.~Pesavento,
\newblock ``{MIMO} radar performance analysis under {K}-distributed clutter,''
\newblock in {\em Proc. ICASSP}, Florence, Italy, May 2014, pp. 5287--5291.

\bibitem{mim6}
K.~Yao,
\newblock ``Spherically invariant random processes: Theory and applications,''
\newblock in {\em Communications, Information and Network Security},
  V.~K.~Bhargava \textit{et al.}, Ed., pp. 315--332. 2002.

\bibitem{mim8}
F.~Gini,
\newblock ``Sub-optimum coherent radar detection in a mixture of
  {K}-distributed and {G}aussian clutter,''
\newblock {\em IEE Proceedings - Radar, Sonar and Navigation}, vol. 114, no. 1,
  pp. 39--48, Feb. 1997.

\bibitem{mim9}
M.~Greco, F.~Bordoni, and F.~Gini,
\newblock ``{X}-band see-clutter nonstationarity: Influence of long waves,''
\newblock {\em IEEE J. Ocean. Eng.}, vol. 29, no. 2, pp. 269--283, Apr. 2004.

\bibitem{mim14}
F.~Pascal, Y.~Chitour, J.-P. Ovarlez, and P.~Forster,
\newblock ``Covariance structure maximum-likelihood estimates in compound
  {G}aussian noise: Existence and algorithm analysis,''
\newblock {\em IEEE Trans. Signal Processing}, vol. 56, no. 1, pp. 34--48, Jan.
  2008.

\bibitem{mim15}
J.~Wang, A.~Dogand\v{z}i\'{c}, and A.~Nehorai,
\newblock ``Maximum likelihood estimation of compound {G}aussian clutter and
  taget parameters,''
\newblock {\em IEEE Trans. Signal Processing}, vol. 54, no. 10, pp. 3884--3897,
  Oct. 2006.

\bibitem{mim17}
Y.~Chitour and F.~Pascal,
\newblock ``Exact maximum-likelihood estimates for {SIRV} covariance matrix:
  Existence and algorithm analysis,''
\newblock {\em IEEE Trans. Signal Processing}, vol. 56, no. 10, pp. 4563--4573,
  Oct. 2008.

\bibitem{lombardo1}
P.~Lombardo and C.~J. Oliver,
\newblock ``Estimation of texture parameters in {K}-distributed clutter,''
\newblock {\em IEE Proceedings - Radar, Sonar and Navigation}, vol. 141, no. 4,
  pp. 196--204, Aug. 1994.

\bibitem{elkorso3}
M.~N. {El~Korso}, A.~Renaux, and P.~Forster,
\newblock ``{CRLB} under {K}-distributed observation with parameterized mean,''
\newblock in {\em Proc. IEEE International Sensor Array and Multichannel Signal
  Processing Workshop (SAM)}, A Coru\~{n}a, Spain, May 2014, pp. 461--464.

\bibitem{ollila1}
E.~Ollila, D.~E. Tyler, V.~Koivunen, and H.~V. Poor,
\newblock ``Complex elliptically symmetric distributions: {S}urvey, new results
  and applications,''
\newblock {\em {IEEE} Trans. Signal Processing}, vol. 60, pp. 5597--5625, Nov.
  2012.

\bibitem{dog1}
A.~Dogand\v{z}i\'{c} and A.~Nehorai,
\newblock ``Generalized multivariate analysis of variance: {A} unified
  framework for signal processing in correlated noise,''
\newblock {\em IEEE Signal Processing Magazine}, vol. 20.

\bibitem{nohara1}
T.~Nohara and S.~Haykin,
\newblock ``Canada east coast trials and the {K}-distribution,''
\newblock {\em Proc. Inst. Electr. Eng. F}, vol. 138, no. 2, pp. 82--88, Apr.
  1991.

\bibitem{lange1}
K.~L. Lange, R.~J.~A. Little, and J.~M.~G. Taylor,
\newblock ``Robust statistical modeling using the t distribution,''
\newblock {\em J. Amer. Stat. Assoc.}, vol. 84, no. 408, pp. 881--896, Dec.
  1989.

\bibitem{jay2}
E.~Jay, J.-P. Ovarlez, D.~Declercq, and P.~Duvaut,
\newblock ``{BORD}: Bayesian optimum radar detector,''
\newblock {\em Signal Processing}, vol. 83, no. 6, pp. 1151--1162, June 2003.

\bibitem{conte2}
E.~Conte, A.~De Maio, and G.~Ricci,
\newblock ``Recursive estimation of the covariance matrix of a
  compound-{G}aussian process and its application to adaptive {CFAR}
  detection,''
\newblock {\em IEEE Trans. Signal Processing}, vol. 50, no. 8, pp. 1908--1915,
  Aug. 2002.

\bibitem{And84}
T.~W. Anderson,
\newblock {\em An Introduction to Multivariate Statistical Analysis},
\newblock Wiley-Interscience, New York, third edition, 2003.

\bibitem{VSO97}
M.~Viberg, P.~Stoica, and B.~Ottersten,
\newblock ``Maximum likelihood array processing in spatially correlated noise
  fields using parameterized signals,''
\newblock {\em {IEEE} Trans. Signal Processing}, vol. 45, no. 4, pp. 996--1004,
  Apr. 1997.

\bibitem{pp95}
A.~Papoulis and S.~U. Pillai,
\newblock {\em Probability, random variables, and stochastic processes},
\newblock McGraw-Hill, New York, 1965.

\end{thebibliography}

\end{document}